\documentclass[twocolumn]{article}
\usepackage[utf8]{inputenc}
\usepackage{graphicx}
\usepackage{geometry}
\usepackage{wrapfig}
\usepackage{hyperref}
\hypersetup{colorlinks=true, linkcolor=blue, filecolor=magenta, urlcolor=cyan}

\usepackage{amsmath}

\newcommand{\apix}[1]{\textsuperscript{#1}}

\begin{document}
\twocolumn[{
\LARGE Exciton lasing in carbon nanotubes below the population inversion threshold via phonon-mediated stimulated emission \\[1.5em]
\large Stefano {Dal Forno}\apix{1}\apix{*},
     Marco Battiato\apix{1}\apix{+}   \\[1em]
\normalsize
\apix{1}School of Physical and Mathematical Sciences, Nanyang Technological University, Singapore \\
\apix{*}tenobaldi@gmail.com \\
\apix{+}marco.battiato@ntu.edu.sg \\
\hrule
\section*{Abstract}
\setlength{\parindent}{10pt}

    \begin{wrapfigure}{r}{0.44\textwidth}
    \includegraphics[width=0.44\textwidth]{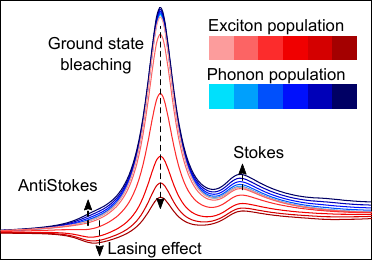}
    \end{wrapfigure}

Excitonic optical transitions in carbon nanotubes (CNTs) have been extensively studied for at least two decades. However, exploiting these transitions to produce a lasing effect has been proved unpractical due to the difficulties of achieving an excitonic population inversion in carbon nanotubes.
In this work we show that lasing is theoretically possible at a much lower exciton population threshold by taking advantage of phonon stimulated emission.

Specifically, the so-called Anti-Stokes transitions, where light is generated by the creation of a phonon and the annihilation of an exciton, can be used to sustain the coherent emission of photons.
These transitions have been overlooked for long due to their low intensity as a consequence of the low occupancy of the associated phonon branches.
We show that the best conditions to achieve the lasing effect are met in CNTs with i) excitons of non--zero momentum, ii) a strong exciton--phonon coupling and iii) narrow exciton transitions.
\\[1em]
KEYWORDS: \textit{carbon nanotubes, excitons, phonons, lasing, anti--Stokes.}
\bigskip\hrule\bigskip}]

\section{Introduction}

Single--wall carbon nanotubes (SWCNTs) are viable systems for a range of electronic and optical applications due to their size and distinctive mechanical, chemical, and electronic properties. The lowest--lying photoexcitations in semiconducting SWCNTs are strongly bound, one--dimensional excitons with binding energies up to 0.5 eV depending on chirality and diameter \cite{wangOpticalResonancesCarbon2005}.
SWCNTs have attracted a lot of attention because of the distinctive optical properties that result from their nearly ideal one--dimensional geometry. 
The mechanisms of light emission in SWCNTs have been widely explored and include light--emitting diodes \cite{misewichElectricallyInducedOptical2003}, photoluminescence spectroscopy \cite{iwamuraNonlinearPhotoluminescenceSpectroscopy2014}, THz emitters and detectors \cite{portnoiTerahertzApplicationsCarbon2008, rosenaudacostaCarbonNanotubesBasis2009}, single-photon sources \cite{hogelePhotonAntibunchingPhotoluminescence2008}, defect--doping \cite{sykesUltrafastExcitonTrapping2019a}, exciton trapping \cite{brozenaControllingOpticalProperties2019}, population inversion of hot electrons \cite{kibisCarbonNanotubesNew2005}, Auger processes \cite{cognetStepwiseQuenchingExciton2007, wangObservationRapidAuger2004} and Raman scattering \cite{jorioRamanSpectroscopyCarbon2021}.

However, an excitonic population inversion effect capable of efficiently sustain coherent light emission in SWCNTs has been not yet achieved. For example, L\"uer et al.~\cite{luerSizeMobilityExcitons2009} demonstrated that with a laser fluence of 4 $\mu$J/cm$^2$ (corresponding to an exciton density of $2 \cdot 10^{13}$ excitons/cm$^2$), only 12\% of the total exciton population phase space is occupied. At this rate, the complete phase space filling would occur at $2 \cdot 10^{14}$ excitons/cm$^2$, but this in fact does not happen because of saturation effects limiting the inversion population process.
This saturation effect is a consequence of scattering processes involving defects, phonons and, most importantly, non--radiative dark excitons. Because of this, the exciton population decays in a strongly dispersive manner with exciton lifetimes that are less than 10 ps~\cite{zhuPumpProbeSpectroscopyExciton2007}. This feature undermines the attempts to achieve an excitonic population inversion.

However, a comparison with other low--dimensional systems suggests that a lasing effect can be achieved with the help of anti--Stokes Raman scattering~\cite{parra-murilloStokesAntiStokesCorrelation2016, hughesPhononmediatedPopulationInversion2013, mansonPolaronMasterEquation2016, quilterPhononAssistedPopulationInversion2015, ferranteRamanSpectroscopyGraphene2018, jorioOpticalPhononResonancesSaddlePoint2014}.
In a \textit{forward} anti--Stokes process light is absorbed, an exciton is created and a phonon is annihilated.
These transitions have always been overlooked due to their low intensity. This is a consequence of the fact that, at room temperature, the population of high-energy phonon is extremely low and this prevents measurements of anti--Stokes transitions in individual nanotubes \cite{gordeevResonantAntiStokesRaman2017}.
On the other hand, a \textit{reverse} anti--Stokes process is also possible: the emission of light by the creation of a phonon and the annihilation of an exciton. This mechanism is at the core of research developments in resonant CNTs optical microcavities~\cite{steinerControllingNonequilibriumPhonon2007}.

In this work we model the absorption and emission of light in CNTs by solving the full quantum Boltzmann equation.
We employ our method to the example case of (6,5) SWCNTs but our conclusions remain general.
We show that coherent light emission in CNTs samples is possible.
We show that this effect can be achieved at a much lower exciton density without the requirement of a complete exciton phase--space filling. This mechanism is driven by the emission of anti-Stokes phonons.
We explore a wide range of initial configurations and determine the optimal conditions under which coherent light emission is achieved.
Our model predicts that light emission is more efficient at i) low temperatures, ii) in the presence of many excitons with non--zero momentum, iii) in cases of strong exciton--phonon coupling and iv) when the main exciton peak is narrower than the phonon side--bands. Our findings are supported by an excellent agreement with pump--probe experiments for low exciton densities.
Most importantly, our model does not use any close--to--equilibrium approximation to describe the exciton and phonon populations~\cite{wadgaonkarNumericalSolverOutofequilibrium2022,waisNumericalSolverTimedependent2021,dalfornoOriginBackgroundAbsorption2022}.

\section{Methods}

A lasing effect is achieved when the process of stimulated emission overcomes the spontaneous emission and absorption of a particular transition.
In practise, determining under which condition we could obtain a net coherent light emission, is equivalent to searching for regions of negative absorbance coefficient.
In this work we model the absorbance of (6,5) SWCNTs in the frequency range around the $E_{11}$ bright excitonic peak.

\subsection{Involved optical transitions}

The time evolution of the population of a system of (quasi)particles can be described by solving the full quantum Boltzmann equation. In the absence of external fields and by neglecting the spatial degrees of freedom this reads
\begin{equation}
    \frac{ \partial f_n}{ \partial t} = \sum_{\alpha} {\left( \frac{ \partial f_n}{ \partial t} \right)}_{\alpha},
    \label{eq:BE}
\end{equation}
where $f_n$ represents the population of a particle labelled by the index $n$, and $\alpha$ runs over all the possible collision terms between such particles.

\begin{figure}[t]
\centering
\includegraphics{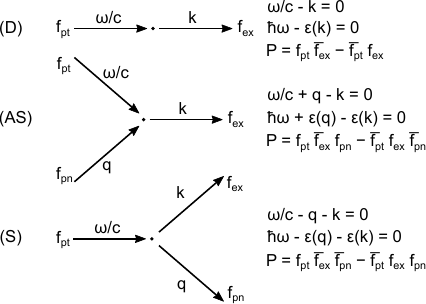}
\caption{Scattering mechanisms in this work.}
\label{fig:1}
\end{figure}

In a similar fashion to our previous work \cite{dalfornoOriginBackgroundAbsorption2022}, we consider three families of collision terms:
\begin{itemize}
    \item the direct absorption (D) of a photon into an exciton;
    \item creation of an exciton via the absorption of a photon and a phonon (Anti--Stokes process, AS);
    \item and the creation of an exciton via the absorption of a photon and the emission of a phonon (Stokes process, S).
\end{itemize}
Figure~\ref{fig:1} shows a pictorial description of the above scattering families and the parameter used in this model are listed in the appendix for the reader's convenience.

We refer to $f_{pt}$, $f_{ex}$ and $f_{pn}$ as the populations of photons, excitons and phonons, respectively.
Also, since we will describe the behaviour of the system out of equilibrium, the populations will not be, in general, a Fermi-Dirac or a Bose-Einstein distribution.

Each transition is constrained by the fermionic or bosonic nature of the quasiparticles and its probability weighted by the population phase--space factor $P$ (see Fig.\ref{fig:1}).  We used the shortened notation $\overline{f_i} = 1 \pm f_i$, where the plus (minus) sign applies for a boson (fermion) quasiparticle.
In this respect, it is important to mention that excitons behave as bosons at very low occupancy, but progressively becomes more fermionic at high populations~\cite{hanamuraCondensationEffectsExcitons1977}.
However, in this work we are doing the simplifying assumption that excitons behave like fermions even for low occupancies.
This approximation holds when Pauli blocking mechanisms start to become important and, as already mentioned above, this happens at a relatively small filling factor~\cite{luerSizeMobilityExcitons2009} (see Supplemental Material for more details \cite{SuppMat}).

Finally, the probability of a specific transition is proportional to the scattering matrix element (see appendix for more details). However the most impactful factor influencing the spectral shape are the  energy and momentum conservation relations, which are again summarised Figure~\ref{fig:1}, where $\hbar \omega / c$, $\hbar k$ and $\hbar q$ are the momenta carried by the photon, exciton and phonon, respectively; while $\hbar \omega$, $\epsilon(k)$ and $\epsilon_{\nu}(q)$ are their energy dispersion relations.

\subsection{Absorption spectrum}

The frequency dependent absorbance $A(\omega)$ is 
\begin{equation}
    A(\omega) \propto \sum_{\alpha} \lambda_{\alpha}(\omega) = \lambda_{D}(\omega) + \lambda_{AS}(\omega) + \lambda_{S}(\omega).
    \label{eq:total-abs}
\end{equation}
In the above equation, $\lambda_{\alpha}(\omega)$ represents the scattering rate of the photon population, i.e.~the \textit{probability rate} of a photon to be absorbed (or emitted) as a consequence of the interaction with the excitons and phonons.
The label $\alpha = \{ D, AS, S\}$ represent one of the three possible scattering events we consider in our model.

The scattering rates $\lambda_{\alpha}(\omega)$ are calculated in terms of Boltzmann integrals.
A full derivation of these integrals is given in Refs.~\cite{waisNumericalSolverTimedependent2021, dalfornoOriginBackgroundAbsorption2022} and it is summarized in the appendix for the reader's convenience.
As an example, we can write the scattering rate of a Stokes process as
\begin{equation}
    \lambda_{S}(\omega) = \frac{2\pi}{\hbar} \frac{1}{V_{BZ}} \sum_{\mathbf{G}}  \int_{V^2_{BZ}} d\mathbf{k} d\mathbf{q} \; |g_S|^2 \; \delta_{\mathbf{k}} \; \delta_{\epsilon} \; P',
    \label{eq:scatt}
\end{equation}
where $V_{BZ}$ is the volume of the Brillouin zone, $|g_S|^2$ is the matrix element of the transition, $\delta_{\mathbf{k}}$ and $\delta_{\epsilon}$ enforce momentum and energy conservation, the summation over $\mathbf{G}$ takes Umklapp scattering into account and $P' = \delta P / \delta f_{pt}$ the derivative of the population factors in Fig~\ref{fig:1}.

\begin{figure}[t]
\centering
\includegraphics{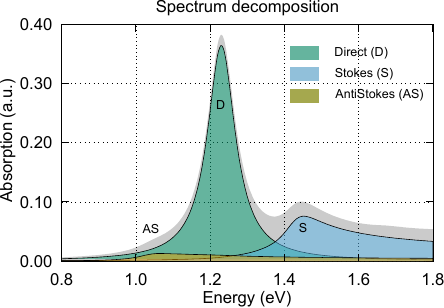}
\caption{Absorption spectrum at equilibrium.}
\label{fig:2}
\end{figure}

Fig.~\ref{fig:2} shows the equilibrium absorption spectrum for (6,5) CNTs decomposed into the D, AS and S components.

\subsection{Approximated lasing thresholds}

It is instructive to write down the integrated peak intensities for the considered processes,
\begin{equation}
    \mathcal{I}_{\alpha} = \hbar \int \lambda_{\alpha}(\omega) d\omega.
    \label{eq:int-peak}
\end{equation}
This quantity is proportional to the total power absorbed by the $\alpha$ process, it can be directly measured in pump--probe experiments and can be used to estimate the density of excitons per unit length as a function of the laser fluence and the number of occupied exciton states~\cite{luerSizeMobilityExcitons2009}.
Within this section we will assume nearly constant matrix elements and $\hbar V_{BZ} \ll 2 m^* c$ (where $m^*$ is the exciton effective mass). Under these approximations, the integrated peak intensity of the mentioned transitions can be evaluated.

In the case of direct absorption we can write
\begin{equation}
	\mathcal{I}_D \sim \frac{2\pi}{\hbar} |g_D|^2 \; \left( 1 - 2 f_{ex}(\Gamma) \right) 
    \label{eq:dep1}
\end{equation}
where $f_{ex}(\Gamma)$ is the exciton population at the $\Gamma$ point, $k=0$. The absorption becomes negative when $f_{ex}(\Gamma) \geq 0.5$.
This condition ensures the presence of lasing when the CNTs are placed in cavity resonant with the associated optical transition and corresponds to the usual requirement of population inversion for any conventional laser. As we have mentioned earlier, it has been shown experimentally that experimentally achievable exciton populations at the bottom of the $E_{11}$ band are far smaller than this threshold, meaning that experimental approaches are still far from reaching excitonic populations that might sustain lasing.

If instead we look at the two phonon-assisted transitions, under the simplifying assumptions mentioned above, we obtain
\begin{align}
    \label{eq:dep2}
    \mathcal{I}_{AS} &\sim \frac{2\pi}{\hbar} \frac{|g_{AS}|^2}{V_{BZ}} \; \left( N_{pn} - N_{ex} - 2 N_{ex} N_{pn}\right) ,\\
    \label{eq:dep3}
    \mathcal{I}_{S} &\sim \frac{2\pi}{\hbar} \frac{|g_{S}|^2}{V_{BZ}} \; \left( 1 + N_{pn} - N_{ex} - 2 N_{ex} N_{pn} \right) ,
\end{align}
where  $N_i = \int d\mathbf{k} f_i(\mathbf{k})$ is the total population of the i-th species. For optical phonons at equilibrium and a sufficiently flat dispersion, $N_{pn}$ is proportional to the Bose--Einstein distribution. Again, we must search for the condition where the integrated peak intensities are negative. We observe that for S transitions, Eq.~\ref{eq:dep3} predicts a very large exciton's populations to achieve a negative absorption. On the other hand, Eq.~\ref{eq:dep2} predicts that, in the case of low temperatures ($N_{pn}\sim0$), \textit{any non--zero exciton population} will produce a negative absorption for AS transitions. At sufficiently low temperatures, the threshold will be far lower that the requirement of population inversion for direct transitions.

While the above result is not completely preserved for more realistic situations, our accurate numerical study will lead to very similar conclusions even when the system is more realistically described.

\subsection{Numerical details}

We compute the above integrals using the TORTOISE package \cite{wadgaonkarNumericalSolverOutofequilibrium2022, waisNumericalSolverTimedependent2021}.
This tool is capable of solving the full BE without any close--to--equilibrium approximations.
The numerical parameters used to describe the system are identical to the ones used in our previous published paper and are listed in the appendix for the reader convenience \cite{dalfornoOriginBackgroundAbsorption2022}.
However, differently from the previous work, in this paper we explore a wide range of initial boundary conditions, with a particular focus on laser realization.

\section{Results and Discussion}

\begin{figure*}[t]
\centering
\includegraphics{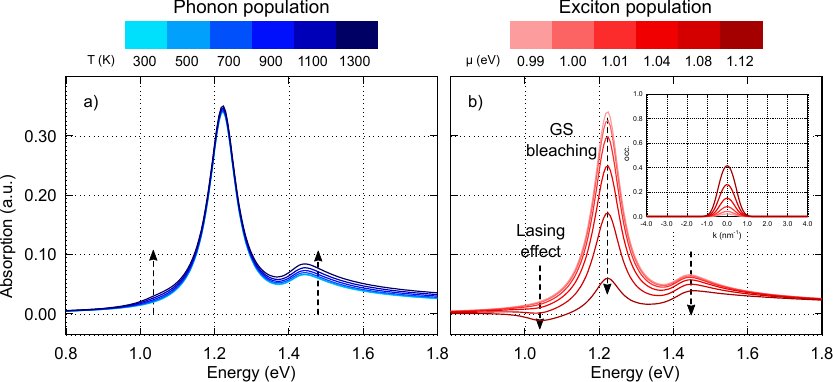}
\caption{Total spectra as a function of photon energy, phonon and exciton populations.}
\label{fig:3}
\end{figure*}

To address the realistic thresholds for lasing in CNTs at different frequencies, we compute the absorption spectra as a function of some representative out--of--equilibrium excitonic populations.
We chose the following test cases
\begin{itemize}
    \item excitons generated optically by a laser,
    \item excitons in thermal equilibrium at a given temperature,
    \item excitons generated in doped CNTs,
    \item excitons in a highly non--thermal configuration.
\end{itemize}
As we will see later in this section, we have observed that only the value of the exciton population at $\Gamma$ and the total number of excitons $N_{ex}$ will play a primary role in our conclusions, while the other details will more marginally affect our results.
For this reason, we choose to show only the absorbance obtained from excitons generated in doped CNTs, and show the remaining ones in the Supplemental Material \cite{SuppMat}.

Fig.~\ref{fig:3} shows the total absorbance obtained in a system of doped CNTs. The inset shows the actual form of the excitonic population and the corresponding chemical potential.
Both plots a) and b) share some similar features: a main prominent peak at $1.24$eV ($E_{11}$ exciton), and two phonon side--bands at $\pm 0.2$eV on the left and right of the main peak.

In plot a) the exciton population has been kept fixed at thermal equilibrium at 300K while the phonon population varies as a function of the temperature. We see that both the Stokes and Anti--Stokes peaks increase as the phonon temperature increases (cf. dashed arrows). This behaviour is confirmed by looking at Eq.~\ref{eq:dep2} and Eq.~\ref{eq:dep3} where the population factors of all transitions depend linearly on $N_{pn}$ at the first leading order. Also, it is important to note the main exciton peak is unaffected by the phonon temperature.

The situation is much different for plot b).
In this case the phonon population has been kept fixed at thermal equilibrium at 300K, while the exciton population varies with respect to the doping level (see inset).
Specifically, we tuned the doping level by changing the exciton chemical potential form $0.99$ eV (undoped) to $1.12$ eV (very doped).
We see that as the exciton population increases, the whole absorption spectrum greatly reduces.
Specifically, we observe two important effects. The weakening of the main excitonic peak at $1.24$eV, also known as ground state (GS) bleaching, and a region of \textit{negative absorbance} close to the anti-Stokes peak at $\sim 1.05$eV, which represents the emission of light by the system.
The first effect, also observed in the work by L\"uer et al.~\cite{luerSizeMobilityExcitons2009}, is caused by a Pauli blocking mechanism that forbids the creation of excitons at the $\Gamma$ point and thus greatly reduce the absorption of light. As previously stated, this effect is maximum (i.e. zero absorbance) when the excitonic filling factor at the $\Gamma$ point is $1/2$. However, as already mentioned, such very high threshold for lasing is currently not experimentally achievable.

The second and most important effect is the emission of light as a consequence of the exciton--phonon interaction.
This effect causes excitons to decay by the emission of a photon and a phonon, see Fig.~\ref{fig:1} (AS). We observe that, as stated earlier in the method section, the lasing threshold of this process is much lower than in the case of the GS bleaching.
We start to observe a negative absorbance at a $\Gamma$ filling factor of roughly $0.26$ and a total exciton density of $0.24\text{nm}^{-1}$ (cf. curve with $1.08$ eV doping).
These densities are lower than the theoretical inversion population threshold, but still very high to achieve experimentally \cite{luerSizeMobilityExcitons2009}.
This is due to the overlap between the main and the AS peaks.
As mentioned previously, the AS peak can become negative for any non--zero exciton population $N_{ex}$. However, the lasing effect is suppressed by the large positive absorption due to the tail of the main exciton peak.
By looking at Fig.~\ref{fig:2} and bearing in mind the lasing conditions from Eqs.~\ref{eq:dep1} and~\ref{eq:dep2}, we see that we need a very high total exciton population $N_{ex}$ so that the AS peak can overcome the absorption of the main peak.
This suggests that, despite the predictions of Eq.~\ref{eq:dep2}, to achieve a lasing effect from the AS transition we need a \textit{narrow main exciton peak} (i.e. long living excitons), a \textit{moderate filling factor} at the $\Gamma$ point and many \textit{excitons with non--zero momentum} (i.e. $N_{ex} > 0$).

\begin{figure*}[t]
\centering
\includegraphics{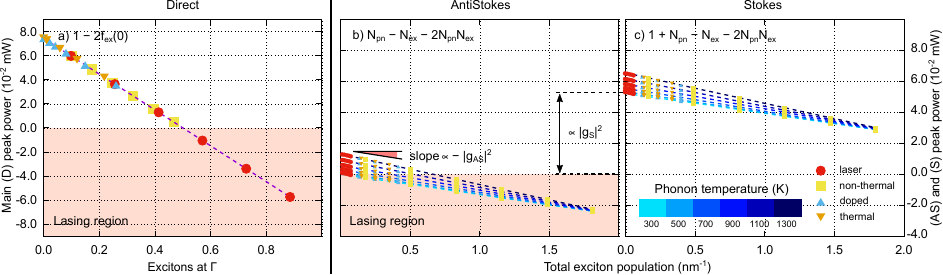}
\caption{D, AS and S peak intensities versus phonon temperature and exciton population.}
\label{fig:4}
\end{figure*}

Now we turn our attention to the integrated peak intensities of the individual transitions.
Fig.~\ref{fig:4} shows the direct, Anti--Stokes and Stokes peak area as a function of the four test cases.
Plot a) shows that the (D) peak decreases linearly with a constant negative slope (cf. Eq.~\ref{eq:dep1}).
Several very important findings can be extracted: i) because of the form of Eq.~\ref{eq:dep1}, the slope is proportional to the value of the matrix element $|g_D|^2$ at the $\Gamma$ point, ii) the point of intersection at $x=0$ equals $\hbar |g_D|^2 / 2\pi$ and iii) the point of population inversion happens at exactly half occupancy $f_{ex}(\Gamma) = 1/2$.
These findings are extremely important because they can be used to better interpret pump--probe experimental results. Specifically, one can estimate the coupling strength $|g_D|^2$ and calculate the number of excited exciton at $\Gamma$ by a linear regression fit.
Note however that these findings have to be considered with care, since we are neglecting second order effects like two--photon excitations and single particle electron--hole excitations.

Similarly, Fig.~\ref{fig:4} b) and c) show the individual dependence of the Stokes and Anti--Stokes integrated intensities as functions of the phonon temperature (blue color scale) and exciton total population (x scale).
The pink shaded region represent the region of negative absorbance, which becomes available only to the AS transition.
We observe two trends: i) the integrated peak intensities decrease as a function of the exciton population and ii) the integrated peak intensities increase as function of the phonon population.
Also, we see that the slope of the AS curve is proportional to the Anti--Stokes coupling strength while the Stokes coupling strength is proportional to the power difference between the AS and S integrated peaks.
These properties tell us that for an efficient lasing effect we need a strong exciton--phonon coupling and an efficient cooling of the sample.

Very interestingly, we also find that this effects depend on the \textit{total} exciton population $N_{ex}$ and does not depend on the actual form of the population function shown in the inset of Fig.~\ref{fig:3} (see also the Supplemental Material \cite{SuppMat}). This means that a broader excitonic population spanning many k-states in the Brillouin zone will produce a stronger lasing effect.
In practise, excitons with non-zero momentum (such as those arising from thermalisation and secondary scattering events) are more likely to decay with the emission of an anti-Stokes phonon and a photon.

Finally, to validate our model, we compare our results to the pump--probe experimental data published by L\"uer et al.~\cite{luerSizeMobilityExcitons2009} (see Supplemental Material for details~\cite{SuppMat, sykesUltrafastExcitonTrapping2019a, luerSizeMobilityExcitons2009, zhengSolutionRedoxChemistry2004}).
\begin{figure}[t]
\centering
\includegraphics{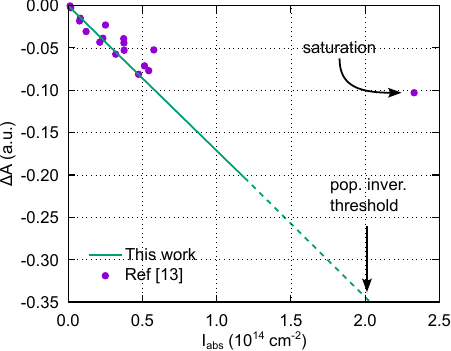}
\caption{Absorbance difference versus pump intensity.}
\label{fig:5}
\end{figure}
Fig.~\ref{fig:5} shows an excellent agreement between our theory and the experimental data.
We observe that for small exciton densities our model correctly reproduce the linear dependence between the exciton population and the D peak intensity. However, at high exciton densities our theory fails to describe the saturation effect visible at the far right side of Fig.~\ref{fig:5} (cf. isolated point).
This has a couple of reasons.
First, the differential absorbance $\Delta A$ is a different quantity than the integrated peak intensity $\mathcal{I}_D$. Although they retain the same behaviour for small laser fluences, they can diverge quite substantially depending on the broadening of the main peak and as a consequence of the overlap with the phonon side-bands. Second, the two quantities $I_{abs}$ and $f_{ex}(\Gamma)$ are linearly dependent only for the ideal case of a Dirac--delta excitonic population centered at $\Gamma$. In real systems, thermalization effects, high laser fluences, exciton--exciton and exciton--phonon scattering broaden the excitonic population.

\section{Conclusion}

We investigated the optical properties of (6,5) SWCNTs and found that a lasing effect can be achieved via the decay of an exciton and the emission of a phonon.
This process requires a much lower lasing activation threshold compared with the main direct excitonic transitions.
However, in the specific case of (6,5) SWCNTs, this threshold is still not low enough to achieve experimentally.
The main factors undermining the lasing effect are the vicinity and the broadening of main excitonic peak.
Thus, for an efficient lasing effect several conditions must be met: i) a narrow main excitonic peak (i.e. long living excitons), ii) a strong exciton--phonon coupling (higher power absorbed by the AS transition), iii) an efficient cooling of the sample and iv) a high population of exciton with non-zero momentum.
The latter can be achieved as a consequence of thermalisation processes or scattering events.
The power emitted is linearly proportional to the total exciton population.

\section*{Acknowledgement}
S.D.F. and M.B. acknowledge the Nanyang Technological University, Singapore, for the support through NAP SUG, and Ministry of Education, Singapore, for the support through MoE AcRF Tier 1 RT13/22, and MOE-T2EP50222-0014.

\section*{Appendix}

We describe the absorption (and emission) of light in (6,5) SWCNTs as a system of interacting excitons, phonons and photons.

\textbf{Band structure and lifetimes.}
The band structure of excitons is approximated by a parabolic dispersion with a given exciton mass and binding energy. In this paper we only model the lowest lying exciton with mass $m_{11} = 0.21 m_e$ and energy $E_{11} = 1.24$ eV, where $m_e$ is the bare electron mass \cite{mattisWhatMassExciton1984, dresselhausExcitonPhotophysicsCarbon2007}.
Among the six phonon branches, we only consider the top four non-acoustic modes, since the former have negligible coupling with excitons \cite{limUltrafastGenerationFundamental2014}.
Also, phonon bands are modelled as nearly flat bands in the vicinity of the scattering regions and their energy are taken from theoretical calculation for consistency \cite{limUltrafastGenerationFundamental2014}.
We label each phonon mode using the standard group theory nomenclature, namely $A$, $E_1$ and $E_2$ phonons \cite{barrosReviewSymmetryrelatedProperties2006}.
Table~\ref{tab:phonon} lists the phonon modes considered in this work.
\begin{table}
\centering
\begin{tabular}{c|ccc}
n-th mode & $A$ & $E_1$ & $E_2$ \\ 
\hline
3rd & 36.5 (RBM) & 50.5 & 76.4 \\ 
4th & 110 (oTO) & 109 & 108 \\
5th & 195 (G--) & 194 & 189 \\
6th & 197 (G+) & 195 & 192
\end{tabular}
\caption{Phonon modes and energies in meV.}
\label{tab:phonon}
\end{table}

Each transition bears an inherent broadening arising from the many--body nature of the interaction. We assign a Lorentzian linewidth to each quasiparticle and use Matthiessen's rule to account for the total broadening of a scattering event.
The values used in this work are shown in Table~\ref{tab:lifetimes} and are taken from theoretical and experimental data \cite{katsutani_direct_2019, kaasbjerg_unraveling_2012, park_electron-phonon_2008}.
\begin{table}
\centering
\begin{tabular}{cccc}
$E_{11}$ & $A^i$ & $E_1^i$ & $E_2^i$ \\
\hline
45.7 & 1.8 & 2.9 & 2.9
\end{tabular}
\caption{Exciton and phonon broadenings in meV.}
\label{tab:lifetimes}
\end{table}

\textbf{Collision integrals and matrix elements.}
In the case of a four leg scattering, the collision integral in the right hand side of Eq.~\ref{eq:BE} can be written as
\begin{equation}
    {\left( \frac{ \partial f_1}{ \partial t} \right)}_{\text{12--34}} = \frac{2\pi}{\hbar} \frac{1}{V^2_{BZ}} \sum_{\mathbf{G}}  \int_{V^3_{BZ}} d\mathbf{k}_2 d\mathbf{k}_3 d\mathbf{k}_4 \; |g_{12}^{34}|^2 \; \,\delta_{\mathbf{k}}\; \delta_{\epsilon}\; P_{12}^{34}.
    \label{eq:collision-term}
\end{equation}
The above equation describes the interaction between two particles in an initial state $|12 \rangle$ that scatter into a final state $|34 \rangle$. Each state is labelled by the wavevector quantum number, $\mathbf{k}_1$, $\mathbf{k}_2$, $\mathbf{k}_3$ and $\mathbf{k}_4$, respectively. Similarly, the individual populations of the initial and final states are $f_1$, $f_2$, $f_3$ and $ f_4$. Eq.~\ref{eq:collision-term}  is a fourth-order integral operator that represents the rate of change of population $f_1$ as a consequence of the interaction with $f_2$, $f_3$ and $f_4$.
We have $g_{12}^{34} = \langle 12| \hat{V}| 34 \rangle = g(\mathbf{k}_1, \mathbf{k}_2, \mathbf{k}_3, \mathbf{k}_4 )$, these are the matrix elements of the interaction $\hat{V}$ between the initial and the final state. In this work, $g_{12}^{34}$ are approximated with averaged constant values obtained from the literature \cite{malicGrapheneCarbonNanotubes2013, jorioCarbonNanotubesAdvanced2008} or fitted to experiments when not available \cite{dalfornoOriginBackgroundAbsorption2022}.
Energy and momentum conservation are enforced by
\begin{equation}
\begin{split}
    \delta_{\mathbf{k}} &= \delta (\mathbf{k}_1 + \mathbf{k}_2 - \mathbf{k}_3 - \mathbf{k}_4 + \mathbf{G}), \\
      \delta_{\epsilon} &= \delta (\epsilon_1(\mathbf{k}_1) + \epsilon_2(\mathbf{k}_2) - \epsilon_3(\mathbf{k}_3) - \epsilon_4(\mathbf{k}_4)),
    \label{eq:dirac-deltas}
\end{split}
\end{equation}
where $\epsilon_i(\mathbf{k}_i)$ are the energy dispersion relations described above.
And finally the population term $P_{12}^{34}$ is given by
\begin{equation}
    P_{12}^{34} = \Bar{f_1} \Bar{f_2} f_3 f_4 - f_1 f_2 \Bar{f_3} \Bar{f_4},
    \label{eq:phase_factor1}
\end{equation}
where $\Bar{f_i} = 1 \pm f_i$, and the plus and minus signs apply to bosonic and fermionic particles respectively.

\textbf{Light absorption.}
The absorption (or emission) of light is computed as the functional derivative of the collision integral with respect to the photon population (cfr. Eq.~\ref{eq:scatt}). For example, given the above four leg scattering, the following quantity
\begin{equation}
    \lambda_{1}(\mathbf{k}_1) = \frac{\delta}{\delta f_1} {\left( \frac{ \partial f_1}{ \partial t} \right)}_{\text{12--34}}
\end{equation}
is the probability rate of change of population $f_1$, as a consequence of the interaction with $f_2$, $f_3$ and $f_4$ \cite{waisNumericalSolverTimedependent2021}.

\bibliographystyle{unsrt}
\bibliography{ms}

\end{document}


\begin{flushleft}

\LARGE
Supplementary Information for:\\
Exciton lasing in carbon nanotubes below the population inversion threshold via phonon-mediated stimulated emission \\[1.5em]

\large
Stefano {Dal Forno}\apix{1}\apix{*},
Marco Battiato\apix{1}\apix{+} \\[1em]

\normalsize
\apix{1}School of Physical and Mathematical Sciences, Nanyang Technological University, Singapore \\
\apix{*}tenobaldi@gmail.com \\
\apix{+}marco.battiato@ntu.edu.sg \\

\end{flushleft}

\hrule\bigskip

\section{Comparison between theory and experiments}

We compare our theoretical calculations with the experimental results obtained by L\"uer et al.~\cite{luerSizeMobilityExcitons2009}.
We use the formula \cite{sykesUltrafastExcitonTrapping2019a, luerSizeMobilityExcitons2009, zhengSolutionRedoxChemistry2004}
\begin{equation}
    I_{abs} = N \frac{A}{\xi^{-1}_C \sigma_C},
    \label{eq:Iabs}
\end{equation}
where $A=0.35$ is the maximum absorbance, $\sigma_C = 7 \cdot 10^{-18} \text{cm}^2$ is the absorption cross section of (6,5) SWCNTs, $\xi^{-1}_C = 4.76 \text{nm}^{-1}$ is the linear carbon atom density and $N$ is the linear exciton density computed from our theory.
Also, we use the same approximation used by L\"uer et al.~and neglect two--photon excitations effects: we assume that one photon excites one exciton. This means that, with a laser fluence of $1\mu J/cm^2$ at 1.24eV, the absorbed pump intensity is $I_{abs} = 5 \cdot 10^{12} \text{photons}/\text{cm}^{2}$.

\section{Exciton statistics}

The properties of the total absorption spectra depends on the choice of the exciton and phonon populations.
We explore the limiting situation where excitons behave as pure Fermions or Bosons.
In real systems, excitons possess a non--trivial statistics that converges to Fermi--Dirac or Bose-Eistein only for very high or very low densities, respectively~\cite{hanamuraCondensationEffectsExcitons1977}.
Table~\ref{tab:1} shows the excitonic population phase factors for Bose and Fermi statistics.
\begin{table}[ht]
\centering
\begin{tabular}{c|cc}
   & Bose exciton    & Fermi exciton \\ 
\hline
D  & 1               & $1 - 2 f_{ex}(\Gamma)$ \\
AS & $N_{pn} - N_{ex}$     & $N_{pn} - N_{ex} - 2 N_{ex} N_{pn}$ \\
S  & $1 + N_{ex} + N_{pn}$ & $1 + N_{pn} - N_{ex} - 2 N_{ex} N_{pn}$
\end{tabular}
\caption{Exciton population phase factors.}
\label{tab:1}
\end{table}

\begin{figure}
\centering
\includegraphics{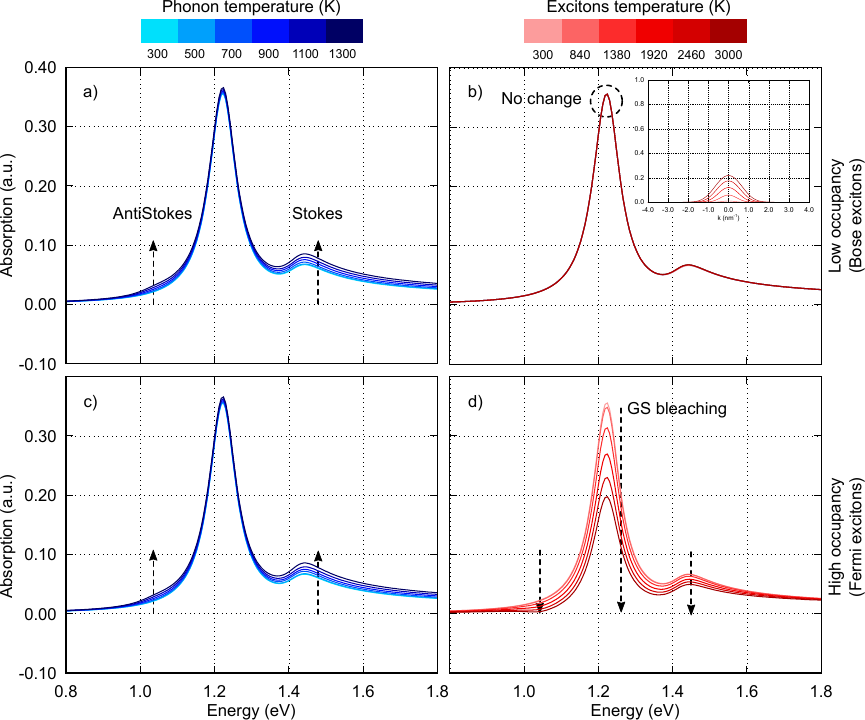}
\caption{Absorption spectra with thermal excitons.}
\label{fig:1}
\end{figure}

\begin{figure}
\centering
\includegraphics{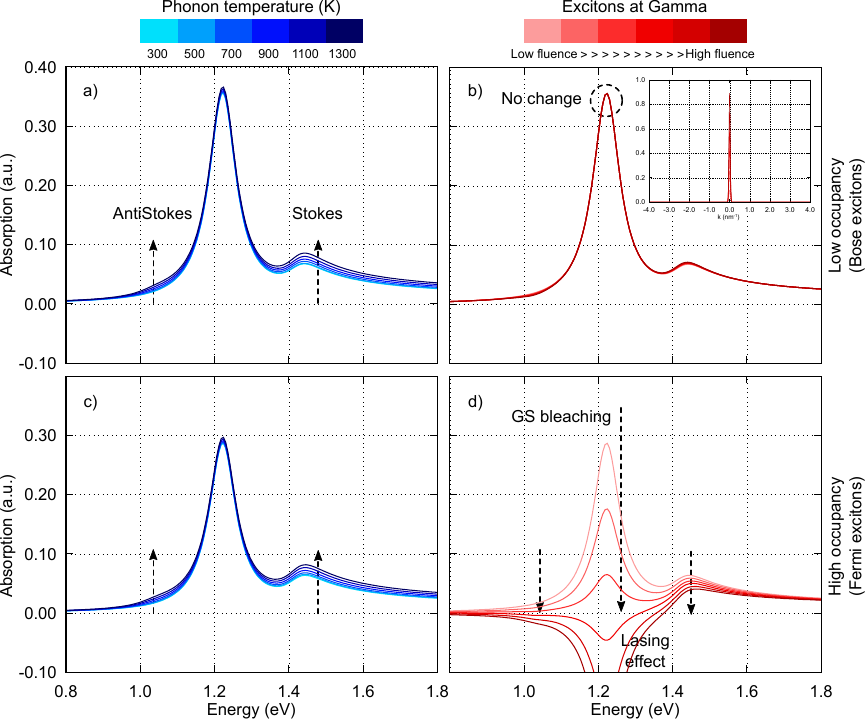}
\caption{Absorption spectra with excitons generated by a laser pulse.}
\label{fig:2}
\end{figure}

\begin{figure}
\centering
\includegraphics{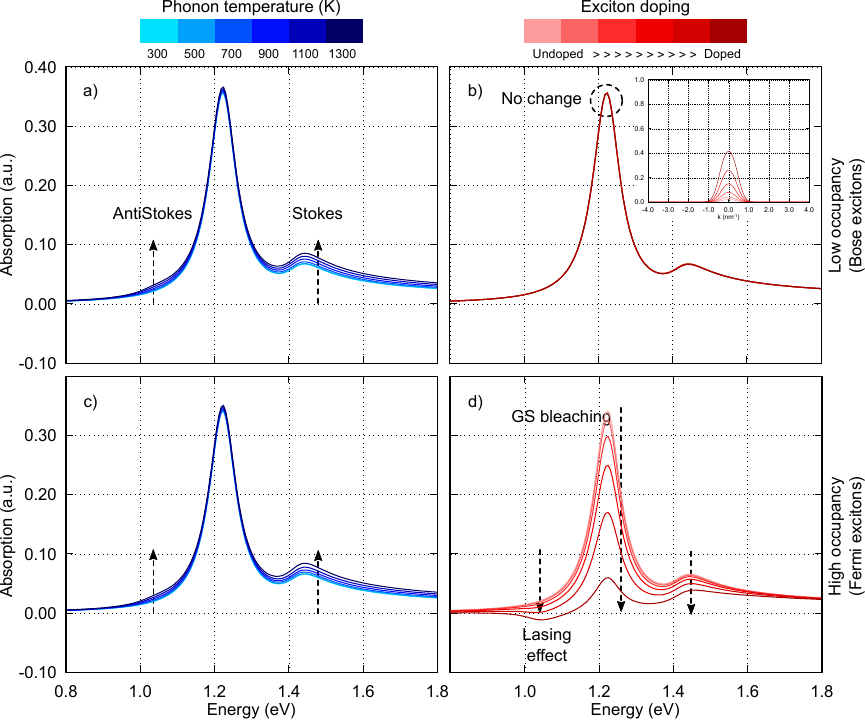}
\caption{Absorption spectra as a consequence of doping.}
\label{fig:3}
\end{figure}

\begin{figure}
\centering
\includegraphics{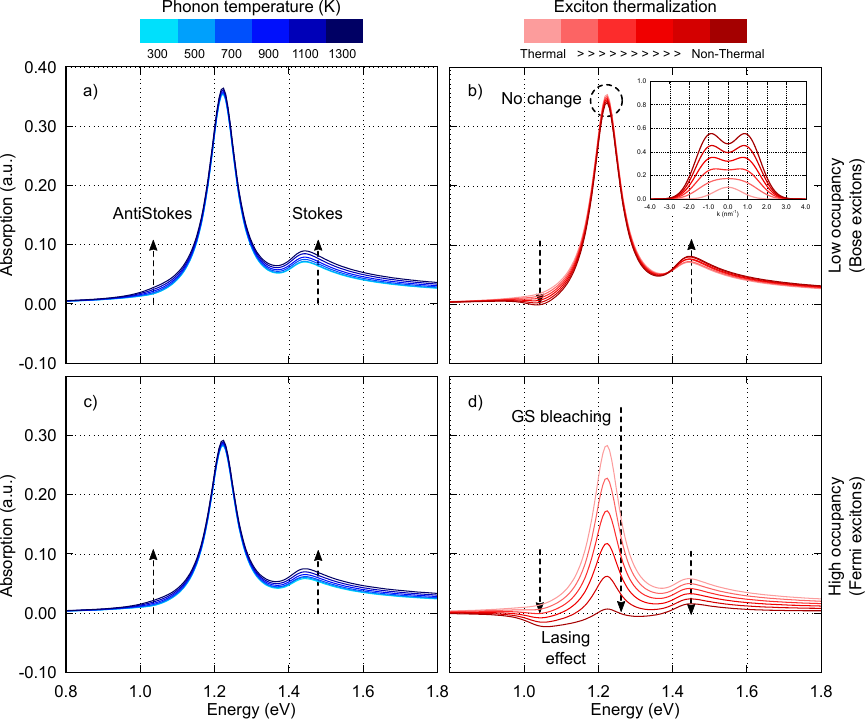}
\caption{Absorption spectra generated by a non--thermal out--of--equilibrium excitonic population.}
\label{fig:4}
\end{figure}

We choose four initial excitonic populations test cases:
\begin{itemize}
    \item an excitonic thermal population, Fig.~\ref{fig:1}.
    \item excitons generated by a laser pulse of given fluence, Fig.~\ref{fig:2}.
    \item excitons created as a consequence of doping, Fig.~\ref{fig:3}.
    \item a non--thermal, out--of--equilibrium population generated by secondary processes, Fig.~\ref{fig:4}.
\end{itemize}
All figures shows very similar trends. All absorption spectra for the Bose excitons show no bleaching effects. This correspond to say that Pauli blocking mechanisms are absent and excitons cannot be modelled as pure bosons.
Similarly, bleaching effect are important in the case of Fermi excitons. Also, when the number of excitons at the $\Gamma$ point is high we achieve the highest bleaching. This is clearly visible in Fig~\ref{fig:2}.
On the contrary, phonon side--bands are unaffected by the number of excitons at the $\Gamma$ point, and instead depend on the total excitonic population $N_{ex}$.

\begin{figure}
\centering
\includegraphics{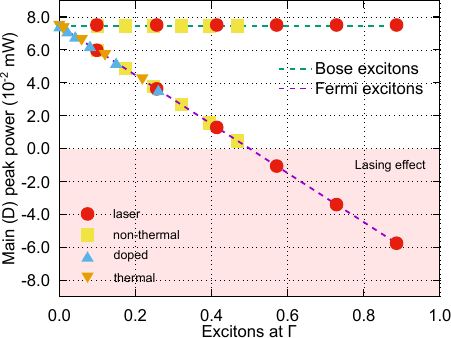}
\caption{Main (D) peak intensity for Bose and Fermi excitons.}
\label{fig:5}
\end{figure}

\begin{figure}
\centering
\includegraphics{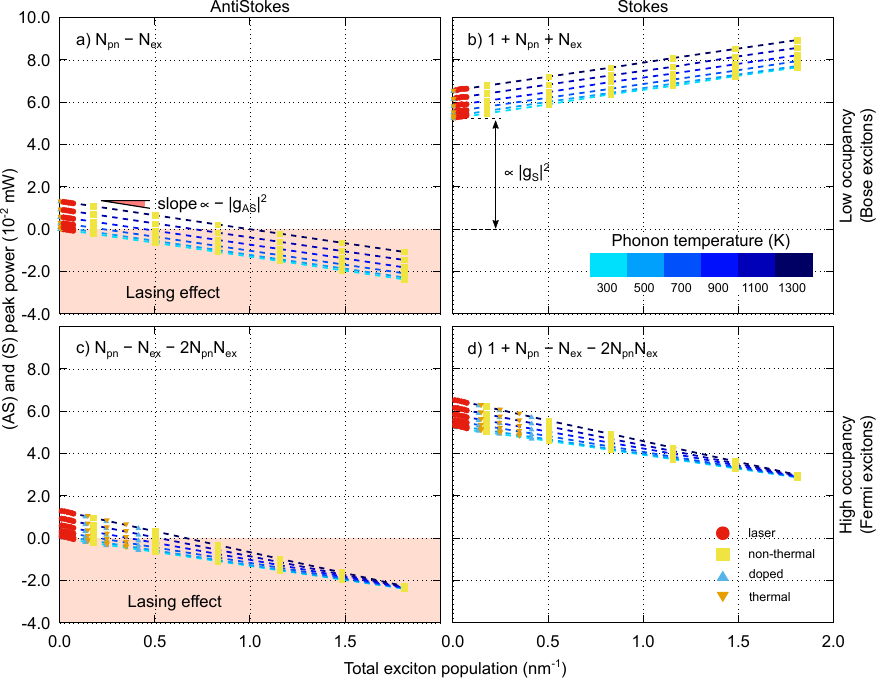}
\caption{(AS) and (S) peak intensities for Bose and Fermi excitons.}
\label{fig:6}
\end{figure}

Finally, Figs.~\ref{fig:5} and \ref{fig:6} show the peak intensities as functions of the exciton population for Bose and Fermi excitons.

\bibliographystyle{plain}
\bibliography{suppinfo}